# Bioregionalization analyses with the `bioregion` R package

Pierre Denelle[1*], Boris Leroy[2+*] & Maxime Lenormand[3+*]

[1] University of Göttingen, Department of Biodiversity, Macroecology & Biogeography, Büsgenweg 1, 37077 Göttingen, Germany

[2] Unité 8067 Biologie des Organismes et Ecosystèmes Aquatiques (BOREA), Muséum national d'Histoire naturelle, Sorbonne Université, Université de Caen Normandie, CNRS, IRD, Université des Antilles, Paris, France

[3] TETIS, Univ Montpellier, AgroParisTech, CIRAD, CNRS, INRAE, Montpellier, France

[*]Corresponding Authors:

pierre.denelle@uni-goettingen.de boris.leroy@mnhn.fr maxime.lenormand@inrae.fr

[+]These authors equally contributed to this work

## Abstract

1. Bioregionalization consists in the identification of spatial units with similar species composition and is a classical approach in the fields of biogeography and macroecology. The recent emergence of global databases, improvements in computational power, and the development of clustering algorithms coming from the network theory have led to several major updates of the bioregionalizations of many taxa.

2. A typical bioregionalization workflow involves five different steps: formatting the input data, computing a (dis)similarity matrix, selecting a bioregionalization algorithm, evaluating the resulting bioregionalization, and mapping and interpreting the bioregions. For most of these steps, there are many options available in the methods and `R` packages.

3. Here, we present `bioregion`, a package that includes all the steps of a bioregionalization workflow under a single architecture, with an exhaustive list of the bioregionalization algorithms used in biogeography and macroecology. These algorithms include (non-)hierarchical algorithms as well as community detection algorithms coming from the network theory. Some key methods from the literature, such as the network community detection algorithm Infomap or OSLOM (Order Statistics Local Optimization Method), that were not available in the R language are included in `bioregion`.

4. By combining different methods coming from different fields to communicate easily, `bioregion` will allow a reproducible and complete comparison of the different bioregionalization methods, which is still missing in the literature.

## 1 | Introduction

Classifying biodiversity into groups sharing similar species composition, i.e. bioregionalization, is an essential aspect of biogeography that allows ecologists to understand how biodiversity assembles at large scales and under what drivers (Lomolino et al., 2017). While bioregionalizations trace back to the early work of Sclater, Darwin and Wallace (Briggs, 2009; Sclater, 1858; Wallace, 1894), the increasing availability of distributional data, the development of new clustering algorithms and improvements in computational power have led to an upsurge in bioregionalization works for many taxa, realms, and geographical areas, such as, to name a few, for continental mammals (Holt et al., 2013; Kreft & Jetz, 2010), birds (Holt et al., 2013; Procheş & Ramdhani, 2012; Rueda et al., 2013), angiosperms (Liu et al., 2023), trees (Droissart et al., 2018), freshwater fish (Leroy et al., 2019), reptiles (Procheş & Ramdhani, 2012), eukaryotic plankton (Sommeria-Klein et al., 2021), deep-sea benthic taxa (Ramiro-Sánchez et al., 2023), marine taxa (Costello et al., 2017), ophiuroids (Victorero et al., 2023). However, there is a large diversity in the methods and approaches used in bioregionalization, especially with recent developments on network-based clustering (Vilhena & Antonelli, 2015), and there is no unifying framework to perform all the steps of a bioregionalization. Here, we present the `bioregion R` package which aims at providing a unique and complete toolbox to perform all the necessary steps of bioregionalization.

A typical bioregionalization analysis consists of four main steps summarized in Figure 1: (1) the formatting of the input data, (2) (facultative) computing (dis-)similarity metric(s) between sites, (3) applying a bioregionalization algorithm and (4) evaluating, ecologically interpreting and mapping the bioregions. At step 1, the formatting of the input data involves switching between matrix and network format depending on algorithm requirements. Step 2 involves computing distance metrics, using up-to-date advances in beta-diversity metrics (Baselga, 2010, 2017) and preferably choosing metrics recommended for bioregionalization (Leprieur & Oikonomou, 2014). Step 3 consists in producing one or several bioregionalizations. A bioregionalization is defined here as the classification of sites (and sometimes species) into bioregions - a bioregionalization is equivalent to a partition in the clustering literature. We divided bioregionalization algorithms into three categories: hierarchical, non-hierarchical, or network-based algorithms. Ecologists have traditionally relied on hierarchical clustering techniques (Kreft & Jetz, 2010), because they produce a dendrogram of bioregions, that supposedly mimics the nested structure of biodiversity (Escalante, 2009; Takhtadzhi͡an & Crovello, 1986). Non-hierarchical algorithms, such as the commonly used K-means algorithm (Hartigan & Wong, 1979) or the Partitioning Around Medoids (PAM; Kaufman & Rousseeuw, 2009), are widely used in other fields and can also be used to generate a predefined set of bioregions. Both hierarchical and non-hierarchical clustering algorithms are based on dissimilarity (or distance) matrices (Figure 1), with few exceptions. More recently, network theory and community detection algorithms have been incorporated into the bioregionalization toolbox (Briega et al., 2023; Denelle et al., 2020; Edler et al., 2017; Lenormand et al., 2019; Leroy et al., 2019; Vilhena & Antonelli, 2015; Yusefi et al., 2019), but remain out of existing bioregionalization packages such as phyloregion (Daru et al., 2020). In some study cases, such algorithms have been shown to produce more relevant results than classical hierarchical clustering (Bloomfield et al., 2018; Vilhena & Antonelli, 2015). Algorithms derived from network theory are mostly similarity-based, but can also work directly on the bipartite network of species occurrences, or abundances, in sites (Beckett, 2016, Figure 1). Step 4 aims at calculating different evaluation metrics to determine the optimal number of bioregions, as well as interpreting and mapping the obtained bioregions.

In the following, we detail the logic of each step of the workflow. For more in-depth use of the package, we have created a dedicated website (due to the double-blind review process, link will be made available after the review; the content of the website is compiled in the supplementary material), with detailed tutorials to guide users through the bioregionalization workflow. The package is available for direct installation via `R` from the Comprehensive R Archive Network (CRAN, https://CRAN.R-

project.org/package=bioregion). The development version is hosted on GitHub (https://github.com/bioRgeo/bioregion).

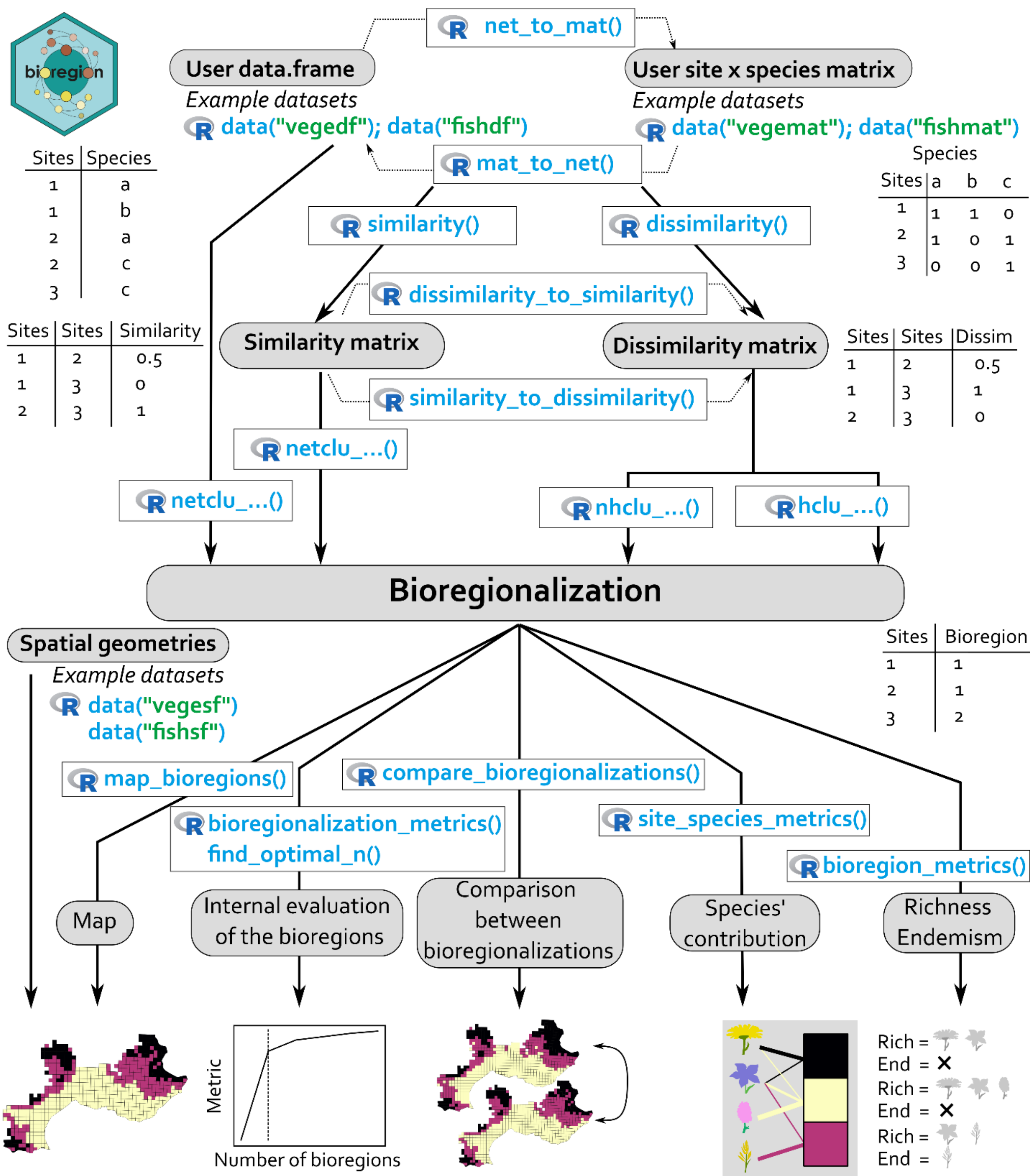

**Figure 1**. Workflow of the bioregion R package.

| Category | Functions | Description | Source package function |
|---|---|---|---|
| **Utils** | `install_binaries()` | Download, unzip, check permission and test the bioregion's binary files | |
| | `mat_to_net()` `net_to_mat()` | Create a data.frame from a site x species matrix and reverse | |
| | `site_species_subset()` | Extract a subset of nodes (sites or species) from a bioregionalization | |
| **Pairwise similarity and distance metrics** | `similarity()` | Compute similarity metrics between sites based on species composition | |
| | `dissimilarity()` | Compute dissimilarity metrics (beta-diversity) between sites based on species composition | |
| | `similarity_to_dissimilarity()` | Convert similarity indices to dissimilarity metrics | |
| | `dissimilarity_to_similarity()` | Convert dissimilarity indices to similarity indices | |
| | `betapart_to_bioregion()` | Convert `betapart` dissimilarity to `bioregion` dissimilarity | |
| **Hierarchical clustering** | `hclu_hierarclust()` | Hierarchical clustering based on dissimilarity or beta-diversity | `fastcluster::hclust()` |
| | `hclu_diana()` | Hierarchical clustering based on dissimilarity or beta-diversity | `cluster::diana()` |
| | `cut_tree()` | Cut a hierarchical tree | |
| | `hclu_optics ()` | OPTICS hierarchical clustering algorithm | `dbscan::optics()` |
| **Non-hierarchical clustering** | `nhclu_dbscan()` | dbscan clustering | `dbscan::dbscan()` |
| | `nhclu_kmeans()` | Non-hierarchical clustering: k-means analysis | `stats::kmeans()` |
| | `nhclu_pam()` | Non-hierarchical clustering: partitioning around medoids | `cluster::pam()` |
| | `nhclu_clara()` | Non-hierarchical clustering: partitioning around medoids | `fastkmedoids::fastclara()` |
| | `nhclu_clarans()` | Non-hierarchical clustering: partitioning around medoids | `fastkmedoids::fastclarans()` |
| | `nhclu_affprop()` | Affinity propagation | `apcluster::apcluster()` |
| **Network clustering** | `netclu_beckett()` | Community structure detection in weighted bipartite network via modularity optimization | `bipartite::computeModules()` |
| | `netclu_infomap()` | Infomap community finding | Code compiled from [www.mapequation.org/infomap/](www.mapequation.org/infomap/) |
| | `netclu_greedy()` | Community structure detection via greedy optimization of modularity | `igraph::cluster_fast_greedy()` |
| | `netclu_labelprop()` | Finding communities based on propagating labels | `igraph::cluster_label_prop()` |
| | `netclu_leadingeigen()` | Finding communities based on leading eigen vector of the community matrix | `igraph::cluster_leading_eigen()` |
| | `netclu_louvain()` | Louvain community finding | `igraph::cluster_louvain()` or code compiled from https://sourceforge.net/projects/louvain/ |
| | `netclu_oslom()` | OSLOM community finding | Code compiled from https://sourceforge.net/projects/louvain/ |
| | `netclu_walktrap()` | Community structure detection via short random walks | `netclu_walktrap()` |
| | `netclu_leiden()` | Finding communities using the Leiden algorithm | `netclu_leiden()` |
| **Clustering analysis** | `bioregionalization_metrics()` | Calculate comparative metrics for entire bioregionalizations | |
| | `find_optimal_n()` | Search for an optimal number of bioregions in a list of bioregionalizations | |
| | `compare_bioregionalizations()` | Compare sites and species memberships across multiple bioregionalizations | |
| | `site_species_metrics()` | Evaluate the contribution of sites and species to bioregions | |

| | `bioregion_metrics()` | Summary statistics for each bioregion | |
|---|---|---|---|
| **Visualisation** | `map_bioregions()` | Map the bioregions | |
| **Datasets** | `vegedf()`<br>`vegemat()`<br>`vegesf()` | Spatial distribution of Mediterranean vegetation (data.frame; co-occurrence matrix and spatial grid) | |
| | `fishdf()`<br>`fishmat()`<br>`fishsf()` | Spatial distribution of fish in Europe (data.frame; co-occurrence matrix and spatial grid) | |

**Table 1**. Overview of the 40 functions available in the `bioregion R` package.

## 2 | Input data

The common use of bioregionalization is to cluster sites based on their species composition, but bioregionalization can span multiple dimensions. Spatially, bioregionalizations range from micro-scale (e.g. gutters of Paris, Hervé et al., 2018) and small-scale studies (e.g. how plots assemble into communities Denelle et al., 2020) to regional and global scales (e.g. ecoregions, biogeographic regions). They can also be applied above (McDonald-Spicer et al., 2019) and below the species level (DiBattista et al., 2017). Furthermore, bioregionalization procedures can be based on either presence-absence or on quantitative measures such as abundance, biomass, area covered or relative abundance, which has implications for the calculation of pairwise dissimilarities between sites and the choice of the clustering algorithms. For presence-absence data, most dissimilarity indices and clustering methods assume absences are reliably recorded.

Second, there are two possible formats for input data. Ecologists are generally accustomed to a site-species matrix – called *matrix format* (Figure 1). This site-species matrix can also be encoded in a 'long-format' data frame, where each species-site occurrence or abundance represents one row of the data frame. This long format can be defined as a *network format* (Figure 1), as it is usually required for network clustering algorithms (Figure 1). More specifically, it is here a bipartite network, because it consists of two types of nodes (species and sites) connected by links representing the occurrence or abundance of species in sites. Site-site and species-species links are not allowed in this type of network. `bioregion` provides the functions `mat_to_net()` and `net_to_mat()` (Table 1) to quickly switch between these two formats.

## 3|(Dis)similarity metrics

Most bioregionalization algorithms require the computation of either similarity or dissimilarity metrics (also called β-diversity). It is important to note that hierarchical and non-hierarchical clustering algorithms mostly work with dissimilarity metrics, whereas network algorithms work with similarity metrics, so the package ships functions to conveniently switch from similarity to dissimilarity (`similarity_to_dissimilarity()` and `dissimilarity_to_similarity()`; Table 1). It is also essential to bear in mind that most network algorithms can take a bipartite species-site network directly as input.

There is a plethora of (dis)similarity metrics available in the ecological literature (Koleff et al., 2003), which span occurrence and abundance-based data. We implemented metrics recommended in the bioregionalization literature such as the turnover metrics (Kreft & Jetz, 2010), but also allow users to define their own turnover formula. Turnover metrics measure the proportion of unique species between pairs of sites (Baselga, 2010). Among turnover metrics, we recommend to use metrics that are not biased by differences in species richness, such as the Simpson dissimilarity index (βsim, Leprieur & Oikonomou, 2014). All these metrics are implemented in `bioregion`.

## 4 | Clustering into bioregions

`bioregion` includes 18 clustering algorithms that can be classified as hierarchical, non-hierarchical, or derived from the network theory (Table 2). To help users understand the different categories of algorithms, we name our function using a prefix, `hclu_`, `nhclu_`, or `netclu_` for hierarchical, non-hierarchical and network-based algorithms respectively, followed by the name of the algorithm. Non-hierarchical and hierarchical clustering require dissimilarity matrices, with the exception of the affinity propagation, while network clustering works with either similarity matrices or bipartite networks. The bioregionalization algorithms available in `bioregion` come from existing R packages, such as `bipartite` (Dormann et al., 2022), `cluster` (Maechler et al., 2023), `fastcluster` (Müllner & Inc, 2021), or `igraph` (Csárdi et al., 2023), as well as external software such as the network algorithms Infomap (Rosvall & Bergstrom, 2008) and the Order Statistics Local Optimization Method (OSLOM, Lancichinetti et al., 2011). All parameters of the original algorithms remain accessible.

| Category | Algorithm | Hierarchical | Dissimilarity or Similarity based |
|---|---|---|---|
| **Hierarchical clustering (bioregions nested in a dendrogram)** | Agglomerative clustering with method Ward.D, Ward.D2, single, complete, UPGMA, WGPMA, WPGMC, UPGMC | Yes | Dissimilarity |
| | Divisive clustering (DIANA) | Yes | Dissimilarity |
| | Ordering Points To Identify the Clustering Structure (OPTICS) (Ankerst et al., 1999) | Yes | Dissimilarity |
| **Non-hierarchical clustering (unique non-nested partition of bioregions)** | Density-based Spatial Clustering of Applications with Noise (DBSCAN) (Ester et al., 1996) | No | Dissimilarity |
| | K-means (Hartigan-Wong, Lloyd, Forgy and MacQueen) (Hartigan & Wong, 1979) | No | Dissimilarity |
| | Partitioning Around Medoids (PAM) (Kaufman & Rousseeuw, 2009) | No | Dissimilarity |
| | Clustering Large Applications (CLARA) (Kaufman & Rousseeuw, 2009) | No | Dissimilarity |
| | Clustering Large Applications based on RANdomized Search (CLARANS) (Kaufman & Rousseeuw, 2009) | No | Dissimilarity |
| | Affinity Propagation (Frey & Dueck, 2007) | No | Similarity |
| **Network clustering (unique non-nested partition of bioregions)** | Newman modularity measure optimized by Beckett (2016) | No | Bipartite network |
| | INFOMAP (Rosvall et al., 2009) | No and Yes | Similarity or bipartite network |
| | Fast greedy modularity optimization (Clauset et al., 2004) | No | Similarity |
| | Propagating labels (Raghavan et al., 2007) | No | Similarity |
| | Leading eigenvector (Newman, 2006) | No | Similarity |
| | Louvain community detection (Blondel et al., 2008) | No | Similarity |
| | Order Statistics Local Optimization Method (OSLOM, Lancichinetti et al., 2011) | No and Yes | Similarity |
| | Random walks (Pons & Latapy, 2005) | No | Similarity |
| | Leiden algorithm (Traag et al., 2019) | No | Similarity |

**Table 2**. Overview of the different algorithms available in `bioregion`. Algorithms can be categorized as hierarchical clustering, non-hierarchical clustering or network clustering.

*Non-hierarchical distance-based clustering*

Classic non-hierarchical clustering algorithms are either centroid-based or density-based. Centroid-based algorithms group sites into the same bioregion if they are close enough to a cluster center. The center of the cluster is determined by optimizing a function that measures the distance between the center and the members of the cluster. We included the two classical algorithms, K-means (Hartigan & Wong, 1979) and PAM (Kaufman & Rousseeuw, 2009) and two algorithms that are less common in biogeography: CLustering LArge Applications (CLARA) and CLARA based on RANdomized Search (CLARANS, Kaufman & Rousseeuw, 2009), both of which are extensions of PAM. We also included one density-based algorithm, the Density-based Spatial Clustering of Applications with Noise (DBSCAN) algorithm (Ester et al., 1996), which identifies areas of high density in the dissimilarity matrix and group them into similar bioregions. Density-based algorithms are rarely utilized in biogeography, with one instance being used to identify endemic plant bioregions in Crete (Kougioumoutzis et al., 2020).

A major methodological aspect of dissimilarity-based non-hierarchical algorithms is their dependence on user-defined number of bioregions. This number is typically found by evaluating multiple bioregionalizations (see Section 5), but they can also be defined *a priori* by users based on their expert knowledge or expectations. A final non-hierarchical algorithm sometimes used in biogeography (e.g., Marchese et al., 2022; Rueda et al., 2013) is the affinity cluster propagation (Frey & Dueck, 2007), which uses the similarity between pairs of sites and searches for clusters in an iterative process. This algorithm does not require an *a priori* number of bioregions.

*Hierarchical clustering*

Hierarchical clustering generally refers to a method of grouping sites in a hierarchical tree (also called dendrogram) on the basis of their dissimilarity. This hierarchical tree can then be cut at different heights to identify bioregions. The hierarchical tree can be constructed in an agglomerative manner (i.e. all sites are initially assigned to their own bioregion and they are progressively grouped together) or in a divisive manner (i.e. all sites belong to the same unique bioregion and they are progressively divided into different bioregions). For divisive hierarchical clustering, we included the DIvisive ANAlysis clustering (DIANA, Kaufman & Rousseeuw, 2009). In agglomerative clustering, a linkage function is used to determine whether two sites should belong to the same bioregion. There are several linkage functions, including the Ward's method, unweighted pair-group method with Arithmetic mean (UPGMA, or between-group average), unweighted pair-group method using centroids (UPGMC), weighted pair-group method using arithmetic averages (WPGMA), weighted pair-group method using centroids (WPGMC), Ward's method, single (SL) and complete linkage (CL). All of these methods are available in the `hclu_hierarclust()` function.

Importantly, the order of sites in the distance matrix can influence the topology of the hierarchical tree, or dendrogram. Indeed, one of the conclusions of Dapporto et al. (2013) was to show that, when tied dissimilarity occurs between pairs of sites during the construction of an agglomerative tree, there is an arbitrary choice of one of the many different trees generated. This arbitrary selection can greatly bias the bioregionalization by producing dendrograms with a low cophenetic correlation coefficient. This coefficient represents the correlation between distances in the distance matrix and the cophenetic distance (i.e., the height of nodes in the hierarchical tree), i.e. how much of the original information is retained in the dendrogram (Sneath & Sokal, 1973).

This major issue is still ignored in most published bioregionalizations based on agglomerative clustering. To tackle this issue, Dapporto et al. (2015) developed a robust method to overcome row-order bias and identify a consensus clustering for a given number of regions (the `recluster.region` algorithm implemented in the `recluster` package (Dapporto et al., 2013)). However, their method did not provide a tree topology, which is generally one of the objectives of hierarchical clustering as it allows understanding the relationships between regions. We address this issue by extending the algorithm of Dapporto et al. (2015) to produce a consensus tree topology. This new algorithm, called Iterative Hierarchical Consensus Tree (IHCT), reconstructs the tree from top to bottom by iteratively identifying the majority-decision binary split among many randomizations. We implemented this algorithm as an argument to `hclu_hierarclust()`, and also included options to select the tree with the highest cophenetic correlation coefficient or to determine a consensus tree (Dapporto et al., 2013). We recommend users to scrutinize the changes in tree topology across randomization trials and understand how well does their final tree represent the distance matrix with the cophenetic correlation coefficient.

We also included the semi-hierarchical clustering approach called Ordering Points To Identify the Clustering Structure (OPTICS, Ankerst et al., 1999). In the OPTICS method, sites are ordered so that the closest sites are neighbors. From this, a 'reachability' distance is calculated between each site. Bioregions are then extracted from this reachability distance in a hierarchical fashion. However, the hierarchical nature of the clusters is not directly provided by the algorithm in a tree-like output, but follows the structure of the 'reachability plot' (Hahsler et al., 2019).

*Network clustering*

Network clustering relies on community detection algorithms, where a community is defined as a cluster of nodes within a larger network - i.e. community in network theory does not convey the same definition as community in ecology. Community detection algorithms aim to identify nodes in a network that are more densely connected to each other than the other parts of the network or than expected by chance (Blondel et al., 2008; Newman, 2006; Traag et al., 2019). To identify such parts of a network, algorithms can maximize a statistic such as modularity – which compares the ratio of within- and between-cluster links (e.g. Blondel et al. (2008), or rely on other principles such as random walks in the network (Pons & Latapy, 2005; Rosvall & Bergstrom, 2008)).

In biogeography, network clustering can be performed using two different approaches, (i) clustering on pairwise site-site similarity matrices, in the form of a weighted unipartite network (e.g., Lenormand et al., 2019), (ii) clustering on bipartite species-site networks, without calculation of similarity metrics (e.g., Leroy et al., 2019; Vilhena & Antonelli, 2015). Bipartite species-site networks can be unweighted (i.e., occurrence of species in sites) or weighted (e.g., abundance, density or biomass of species in sites).

We implemented the most common community detection algorithms in `bioregion` in functions with the prefix `netclu_` followed by the name of the algorithm, e.g. `netclu_infomap`. Among these algorithms, most work with both unipartite (argument `bipartite = FALSE`) and bipartite approaches (argument `bipartite = TRUE`). Such algorithms include Infomap (Rosvall et al., 2009), walktrap (Pons & Latapy, 2005), fastgreedy (Clauset et al., 2004), label propagation (Raghavan et al., 2007), leading eigenvector (Newman, 2006); OSLOM (Lancichinetti et al., 2011), Leiden (Traag et al., 2019) and the Louvain algorithm (Blondel et al., 2008). We also included two algorithms that only work on weighted bipartite networks: Label Propagation Algorithm based on weighted bipartite modularity optimization (LPAwb+) for large networks and its thorough search extension DIRTLPAwb+ for small networks, Beckett (2016).

Two important algorithms, OSLOM (Lancichinetti et al., 2011) and Infomap (Rosvall et al., 2009), were not available in the `R` language so far. We compiled and implemented them in `bioregion`. To use them, the function `install_binaries()` has to be run first. It will install the binary files needed to run both algorithms.

## 5 | Evaluation of the bioregions

Clustering evaluation is the process of evaluating the quality of bioregions in terms of how well the algorithm has grouped similar sites and species together and separated dissimilar ones. Clustering evaluation has multiple dimensions: (1) evaluating the clustering structure with bioregionalization metrics (referred to as 'internal evaluation' in the clustering literature), (2) visual inspection of bioregion results and, when applicable, hierarchical relationships among bioregions, (3) comparing similarity among bioregionalizations or against an external validation bioregionalization (referred to as 'external evaluation' in the clustering literature) and (4) assessing the contribution of sites and species to bioregions. The package comes with functions for all four dimensions of clustering evaluation (Table 1).

*Evaluating the bioregionalization structure and choosing an adequate number of bioregions*

Evaluation metrics are statistics that summarize the quality of the bioregionalization. Generally, they are used for those algorithms that require a predefined number of bioregions, to help users identify an adequate number. However, they can also be used to make quantitative comparisons between different algorithms. Many evaluation metrics exist, but not all metrics make sense to evaluate bioregionalization results. We have included four metrics used or recommended in biogeography, all of which can be computed with the `bioregionalization_metrics()` function:

- The percentage of dissimilarity explained by bioregions which is the sum of within-cluster β-diversity divided by the total β-diversity of the distance matrix (Holt et al., 2013).
- The ANalysis Of SIMilarity (anosim) statistic (Clarke, 1993), which compares the between-cluster dissimilarities to the within-cluster dissimilarities.
- The average endemism among bioregions (Kreft & Jetz, 2010)
- The total endemism across all bioregions (Kreft & Jetz, 2010)

For algorithms requiring a user-defined number of bioregions, we provide guidelines based on standard practices in biogeography in the online help on how to identify an appropriate number of clusters with the `find_optimal_n()` function, which involves calculating metrics for a range of possible numbers of bioregions, and inspecting evaluation plots (evaluation metric plotted against the number of bioregions). Unlike existing bioregionalization packages (Daru et al., 2020), `find_optimal_n()` provides the user with several methods to determine an optimal number of bioregions. However, we recommend biogeographers to question the assumption that there is only one optimal number of clusters, which may be an oversimplification of the biological reality when a hierarchy of bioregions is expected. For example, Ficetola et al. (2017) investigated shallow, intermediate and deep bioregions, using different levels of dissimilarity, which provides a more complete description of the hierarchy of natural regions.

*Mapping bioregions for visual inspection*

Mapping the bioregions obtained in a bioregionalization analysis is important to assess the quality of bioregions (e.g., spatial distribution, cohesiveness of regions; Bloomfield et al., 2018; Victorero et al., 2023) and to understand and test the environmental drivers of a bioregionalization (Daru et al., 2020; Ficetola et al., 2017; Lomolino et al., 2017). If sites, or the spatial unit used, come with a ready to use shapefile, we can then link the obtained bioregions with its spatial component and map the bioregions with the function `map_bioregions()`, which relies on the `sf` package (Pebesma et al., 2023, Figure 2).

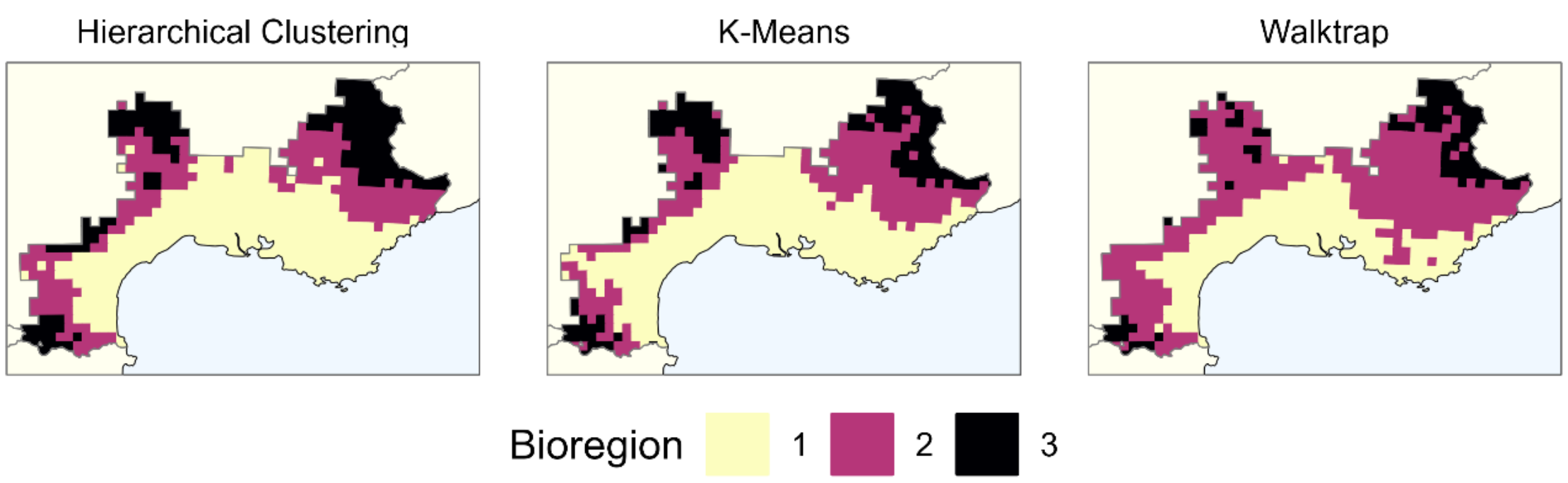

**Figure 2**. Spatial mapping of three different bioregionalizations obtained using the hierarchical clustering with McQuitty linkage, K-means (Hartigan & Wong, 1979), and network walktrap approaches (Pons & Latapy, 2005), applied to the vegetation dataset provided in `bioregion`. The `hclu_hierarclust()`, `nhclu_kmeans()` and `netclu_walktrap()` functions from the package were used to generate this figure. The R code needed to reproduce this figure is available in the Appendix.

*Comparing different bioregionalizations*

Comparing different bioregionalizations is useful for i) comparing the results with an external validation, ii) identifying which algorithms produced similar results, iii) assessing the uncertainty of stochastic algorithms such as Infomap, where bioregionalization results can be highly variable between runs (e.g., Brown et al., 2023). `compare_bioregionalizations()` allows to do so using two metrics based on the confusion matrix between pairs of bioregionalizations, the Rand index (Rand, 1971) and the Jaccard index.

*Summary statistics*

Computing individual summary statistics per site, species, and bioregion is a necessary step to understand the composition and nature of bioregions. For sites and species, we have implemented the function `site_species_metrics()`, which computes metrics that quantify the contribution of species or sites to each bioregion. The metrics include the participation coefficient *C* and the relative intra-module degree *z* described in Guimera & Nunes Amaral (2005), the contribution metric $\rho$ used in Lenormand et al. (2019) as well as the fidelity and affinity metrics used in Bernardo-Madrid et al. (2019). For bioregion-level metrics, we implemented the `bioregion_metrics()`, which provides measures of richness, endemism, and spatial coherence of bioregions, as defined in Divíšek et al. (2016).

## Conclusion

Because of biogeographical connections, environmental gradients, and biological processes such as speciation and biotic dispersal, bioregions are usually considered to be hierarchically ordered (Kreft & Jetz, 2010; McLaughlin, 1992). This has led to the popularity of hierarchical clustering algorithms in bioregionalization literature. The emergence of community detection algorithms coming from the network theory led to important studies with updated maps of natural bioregions (Edler et al., 2017; Leroy et al., 2019; Rojas et al., 2017; Vilhena & Antonelli, 2015) or the recent exploration of Anthropocene bioregions (Bernardo-Madrid et al., 2019; Brown et al., 2023; Capinha et al., 2015; Leroy et al., 2023). However, with a few exceptions (Bloomfield et al., 2018; Hill et al., 2020), no large-scale study has yet compared the different methods. By putting all the different steps of the bioregionalization workflow in a common place with the same architecture, `bioregion` provides the perfect background to realize such a study.

## Acknowledgments

We thank Raphaëlle Gladieu, Valentin Dijan and Maïri Souza Oliveira and other beta-testers of the `R` package for providing useful insights. BL and ML were funded by their salaries as French public servants.

## Conflict of interest statement

The authors have no conflict of interest to declare.

## Author's contributions

Pierre Denelle, Boris Leroy and Maxime Lenormand developed the `bioregion` `R` package and the associated website. Pierre Denelle led the writing of the manuscript. All authors wrote the tutorials on the associated website, contributed critically to the drafts and gave final approval for publication.

## Data accessibility

The `bioregion` `R` package is available for download from CRAN at https://CRAN.R-project.org/package=bioregion. The development version of the package is available at https://github.com/bioRgeo/bioregion. An associated website with tutorials is available at https://biorgeo.github.io/bioregion/. Updates to the package are submitted to the CRAN repository and archived on the Zenodo repository, with each new release having its own DOI (version 1.2 of the package is archived at https://doi.org/10.5281/zenodo.14781826, Denelle et al., 2024).

```r
# 1. Input data --------------------------------
# install.packages("bioregion")
library("bioregion")
data("vegedf"); data("vegemat")
# 2. (dis)Similarity ---------------------------
d <- dissimilarity(vegemat)
s <- similarity(vegemat)
# 3. Bioregionalization ------------------------
bio1 <- nhclu_kmeans(d, n_clust = 3)
bio2 <- hclu_hierarclust(d, n_clust = 3)
bio3 <- netclu_walktrap(s)
# 4.1. Evaluation ------------------------------
bio2_opt <- hclu_hierarclust(d, n_clust = 2:100)
eval_tree <-
bioregionalization_metrics(bio2_opt, d)
find_optimal_n(eval_tree)
# 4.2. Mapping ---------------------------------
data("vegesf")
map_bioregions(bio1, geometry = vegesf)
# 4.3. Comparison ------------------------------
comp <- merge(bio1$clusters, bio2$clusters,
              by = "ID")
colnames(comp) <- c("ID", "hclu", "nhclu")
comp_part <-
compare_bioregionalizations(comp[, c(2, 3)])
# 4.4. Summary statistics-----------------------
sp_rho <- site_species_metrics(bio1, vegemat,
indices = "rho")
bioregion_stat <- bioregion_metrics(bio1,
 vegemat, vegemap)
```

**Box 1.** A minimalistic representation of the R code for implementing the bioregionalization workflow.